\documentstyle[11pt,ysc,twoside,epsf]{article}

\def\edcomment#1{\iffalse\marginpar{\raggedright\sl#1\/}\else\relax\fi}

\reversemarginpar

\begin{document}

\title{Diffuse Interstellar Bands and Their Families }

\author{ Bogdan Wszo\l ek }

\affil{ Jan D\l ugosz Academy, Insitute of Physics,
Cz\c{e}stochowa, Al. Armi Krajowej 13/15, 42-200 Cz\c{e}stochowa,
Poland. \\ E-mail:bogdan{@}ajd.czest.pl}

\begin{abstract}

Diffuse interstellar bands (DIBs) still await an explanation. One
expects that some progress in this field will be possible when all
the known DIBs are divided into families in such a way that only one
carrier is responsible for all bands belonging to the given family.
Analyzing high resolution optical spectra of reddened stars we try
to find out spectroscopic families for two prominent DIBs, at 5780
and 5797 angstroms. Among the DIBs, observed in the spectral range
from 5590 to 6830 angstroms, we have found 8 candidates to belong to
5780 spectroscopic family and the other 12 DIBs candidating to
family of 5797 structure.

\end{abstract}

\section{Introduction}

The DIBs are absorption features which are generated in the
interstellar medium by still unidentified set of carriers. They are
found in the visual and near infrared spectra, between 4000 and
13500 angstroms. The discovery of the first DIBs in stellar spectra
dates back to the pioneering years of stellar spectroscopy. The
original report on the discovery of two spectral features, centered
near 5780 and 5797 angstroms, in spectra of some spectroscopic
binaries was published in 1922 by Heger. The extended review paper
presenting DIB problematics was published by Herbig (1995).

To date more than 300 DIBs are detected and the number is still
increasing (see e.g. Galazutdinov at all., 2000). None of them has
been identified! The identification of the carriers of DIBs is one
of the most difficult challenges for spectroscopists. The solution
of the mystery of the carriers of DIBs is expected from
interdisciplinary spectroscopic collaboration between molecular
physicists, molecular chemists and astronomers.

\section{DIBs' observing strategy}

To detect DIBs in stellar spectra we need:

(i) relatively bright star of an early spectral type (because in
optical spectra of such stars we have relatively smooth continuum
with only few strong stellar lines, and therefore DIBs are not lost
in the "forest" of stellar lines),

(ii) transparent interstellar diffuse cloud(s) in line of sight
causing reddening of the target star,

(iii) rather big telescope and spectrograph of a very good quality,
to give spectra of sufficiently high resolution and of high signal
to noise ratio.

To be sure that one has to deal with DIB, and not e.g. with any weak
stellar line, the spectroscopic binaries are usually in use. When
one compares few spectra of spectroscopic binary (e.g. spectra
registered in subsequent nights), one can notice that some lines
changed their positions (stellar lines) contrary to the other ones
which are always at the same place.

\section{Observing constraints on DIBs' problem}

The most recent high-quality observational data and the parallel
theoretical studies allow us to define some constraints on the DIBs'
problem:

(i) the presence of DIBs is related to the colour excess, in the
sense that the lack of reddening implies the absence of the DIBs;
but the DIBs' intensity along one line of sight is only loosely
correlated to the value of the reddening in that direction,

(ii) the DIBs' intensity is not strongly correlated with the 2200
angstroms extinction bump: dust and carriers of DIBs, although
coexisting in the interstellar medium, have an independent history,

(iii) the line profile of the DIBs seem to be quite stable,

(iv) the DIBs seem to be generated not by a single agent but by
several carriers; they can therefore be grouped into families, the
members of which show well-defined intensity ratios (e.g., 5780 and
5797 DIBs have different carriers because the ratio of their
intensities may substantially vary from one line of sight to the
other).

All known DIBs form very inhomogeneous sample. Some of them are
relatively strong, contrary to the others which are extremely weak.
There are DIBs which are narrow, with widths of about 1 angstrom
(e.g. 5797), but there are also very broad DIBs, with widths of
about 20 angstroms (e.g. 4430).

\section{The problem of carriers}

Carriers of DIBs, which are still not found, are probably large
interstellar molecules in gas or solid phase. Some authors argue
that Polycyclic Aromatic Hydrocarbons, carbon chains or fullerenes
may be carriers of some DIBs. There are however more exotic
expectations which involve even extremely large organic molecules
which would be responsible for spreading of life in the universe.

\section{Spectroscopic families of DIBs}

It is expected that a progress will be possible, and some carriers
will be closer to be identified, when all known DIBs are divided
into spectroscopic families in such a way that only one carrier is
responsible for all bands belonging to a given family. The dividing
of such kind is to do only by analysis of astronomical spectra.

Few years ago we published a few promising methods to isolate
spectroscopic families of DIBs (Wszo\l ek \& God\l owski, 2003). One
of proposed methods is a direct visual investigating of the figures
with arranged spectrograms of different stars. Using this method one
can easily find DIB candidates which tend to be members of the same
spectroscopic family.

\section{Looking for spectroscopic relatives of 5780 and 5797 DIBs}

Among all known DIBs, two lines are special. These are 5780 and
5797 DIBs. They are the first two DIBs discovered by Heger about
85 years ago. They are positioned very close one to the other in
the spectra. Both lines are relatively strong and narrow. The
intensities of these bands are quite good correlated when few
dozens of lines of sight is taken into account, but there are few
examples showing that intensity ratio for these bands can change
itself very much when going from one target star to another (Kre\l
owski \& Westerlund, 1988). That means that 5780 and 5797 do not
belong to the same spectroscopic family. These DIBs may play a
role of the "heads" of their own spectroscopic families of DIBs.
It is interesting to find spectroscopic "relatives" for each of
these two bands, and this was a subject of our last investigation.

We used spectra registered in 1993 in McDonald Observatory (2.08 m
telescope, echelle spectrograph) by Professor Jacek Kre/l owski.
They cover spectral range from 5590 to 6830 angstroms, in which
almost 200 of known DIBs are present.

Looking for spectral "relatives" of 5780 band we used spectra of
three stars (HD): 23180, 166937 and 206165. We arranged spectra of
the stars in such a way that intensity of DIB 5780 was rising as one
followed from one spectrum to the other. In such arranging,
intensities of these bands which belong to 5780 spectroscopic family
should exactly follow the change of intensity of 5780. Visual
comparison of arranged spectrograms, taken for all rows of spectra,
allowed us to indicate some member candidates to 5780 spectroscopic
family. The most promising candidates are DIBs positioned at 5776,
5795, 6108, 6162, 6697, 6793, 6795 and 6827 angstroms.

The similar analysis but using three other stars, was carried out
for DIB at 5797 angstroms. This time, target stars were chosen in
such a way that they all give almost equal intensities of 5780 band
while the intensity of 5797 is changing very much. These stars are
(HD): 147165, 206267 and 207198. DIBs which tend to follow intensity
change of 5797 line, and therefore they candidate to 5797
spectroscopic family are these positioned at: 5719, 5766, 5769,
5773, 5793, 5819, 5829, 5850, 6090, 6439, 6449 and 6492 angstroms.
The best candidates are bands at 5793, 5829 and 5850 angstroms.

Further investigation, based on better data samples and involving
other spectral ranges, is necessary to confirm above candidates and
to complete spectroscopic families of 5780 and 5797 DIBs. Also,
looking for the further "head" bands, like 5780 and 5797 ones, is an
important line of investigation and it is in progress.

\section{Conclusion}

The problem of DIBs' carriers, due to its very interdisciplinary
character, is a very promising subject for open minded young
scientists. Although it is difficult to be solved fast, researcher
who investigates DIBs has many occasions to feel a thrill of
discovery.

\begin {references}

\reference Galazutdinov G.A., Musaev F.A., Kre\l owski J., Walker
G.A.H.: 2000, PASP, 112, p.648;

\reference Heger M.L.: 1922, Lick. Obs. Bull., 10, p.146;

\reference Herbig G.H.: 1995, ARA\&A, 33, p.19;

\reference Kre\l owski J., Westerlund B.E.: 1988, A\&A, 190,
p.339;

\reference Wszo\l ek B., God\l owski W.: 2003, MNRAS, 338, p.990;

\end {references}

\end{document}